# Chiral nematic self-assembly of minimally surface damaged chitin nanofibrils and its load bearing functions

*Dongyeop X. Oh[1,+], Yun Jeong Cha[2,+], Hoang-Linh Nguyen[3], Hwa Heon Je[3], Yong Seok Jho[4,5], Dong Soo Hwang[3,6,\*], & Dong Ki Yoon[2,\*]*

[1]Research Center for Industrial Chemical Biotechnology, Korea Research Institute of Chemical Technology (KRICT), Ulsan 44429, Republic of Korea

[2]Graduate School of Nanoscience and Technology and KINC, KAIST, Daejeon 305-701, Republic of Korea

[3]Division of Integrative Biosciences and Biotechnology, Pohang University of Science and Technology (POSTECH), Pohang 790-784, Republic of Korea

[4]Department of Physics, Pohang University of Science and Technology, [5]Asia Pacific Center for Theoretical Physics, Pohang 790-784, Republic of Korea

[6]School of Environmental Science and Engineering, Pohang University of Science and Technology (POSTECH), Pohang 790-784, Republic of Korea

[+]These authors contributed equally to this work.

*Correspondence should be addressed to Dong Soo Hwang (e-mail: dshwang@postech.ac.kr) & Dong Ki Yoon. (e-mail: nandk@kaist.ac.kr)

*Abstract*

Chitin is one of the most abundant biomaterials in nature, with $10^{10}$ tons produced annually as hierarchically organized nanofibril fillers to reinforce the exoskeletons of arthropods. This green and cheap biomaterial has attracted great attention due to its potential application to reinforce biomedical materials. Despite that, its practical use is limited since the extraction of chitin nanofibrils requires surface modification involving harsh chemical treatments, leading to difficulties in reproducing their natural prototypal hierarchical structure, i.e. chiral nematic phase. Here, we develop a chemical etching-free approach using calcium ions, called "natural way", to disintegrate the chitin nanofibrils while keeping the essential moiety for the self-assembly, ultimately resulting in the reproduction of chitin's natural chiral structure in a polymeric matrix. This chiral chitin nanostructure exceptionally toughens the composite. Our resultant chiral nematic phase of chitin materials can contribute to the understanding and use of the reinforcing strategy in nature.

*Introduction*

Natural hard tissues such as bones and exoskeleton are composed of organic/inorganic hybrid composites, in which the hierarchically assembled one-dimensional fillers are impregnated to provide strong mechanical properties[1-5]. The exoskeletons of arthropods are built from chitin nanofibrils embedded in calcium carbonate or/and protein matrices (Fig. 1a)[6-14]. The chitin nanofibrils (Fig. 1aII), which consist of rigid crystalline and flexible amorphous regions (Fig. 1aIII, IV), stack together to form the layered structures[15]. These structures form the basic units of the twisted plywood structure (Fig. 1aI), which is a chiral nematic phase and critical for the strong load-bearing property of the exoskeleton[5,11-14]. This property has encouraged studies on the chiral nematic shape of chitin, with the objective of reinforcing biocompatible nanocomposites. However, the harsh chemical treatment necessary to effectively detach the individual building blocks from chitin materials hinders the reproduction of the chiral nematic phase and hence its application.

In the absence of calcium carbonates or proteins, chitin nanofibrils strongly agglomerate, not in the exploitable nanocomposite filler form, but rather in a meaningless lump due to the strong interfibrillar hydrogen bonding (H-bond) (Fig. 1b)[16]. The two most well-known chitin nanomaterials obtained using the harsh treatments are chitin nanowhiskers[9,17] and chitin nanofibers (Fig. 2a,b)[18,19]. Both nanomaterials show the no long-range ordered optical textures in polarized optical microscopy (POM) images (Fig. 2c,f). In enlarged views using scanning electron microscopy (SEM) and transmission electron microscopy (TEM), chitin nanowhiskers and nanofibers are randomly oriented (Fig. 2d, e, g and h).

Mostly, a rich amount of multivalent metals such as calcium (Ca), zinc (Zn), and iron (Fe) are found in most load-bearing chitin-based exoskeletons[6,8,12]. Although the relationship between the incorporated metals and the mechanical properties is not clearly understood, it has been reported that chitin strongly binds the metals[6,12,20]. Recently, the Tokura group reported that Ca-saturated methanol dissolves chitin at molecular level under mild conditions without strong base or acid[10]; it also shows the interaction between chitin and $Ca^{2+}$ ion. If $Ca^{2+}$ ions fully surround the chitin nanofibrils, the interfibrillar H-bonds could be screened.

Based on the binding between $Ca^{2+}$ and the hydroxyl groups of chitin[20,21], we developed a new strategy 1) to physically disintegrate chitin to a nanoscale 1D structure called "chitin nanowire" using our modified Ca-saturated methanol (Ca-methanol) and 2) to control 3D hierarchical structure of the chitin nanowires (Fig. 1b, Fig. 3a-c and Fig. S1). Empirically, the binding affinity of the methanol-solvated $Ca^{2+}$ in the Ca-methanol to the chitin nanowire is comparable with the H-bond between chitin nanowires and the probability to replace the densely cross-linked H-bonds inside the chitin crystalline structure is extremely low due to the limited diffusion of methanol-solvated $Ca^{2+}$. However, the interfibrillar region is permeable to methanol-solvated $Ca^{2+}$, leading to the dispersion of the chitin nanowires without damage to the acetyl groups if the concentration of $Ca^{2+}$ is sufficient to partially replace the interfibrillar H-bond. After the disassembly process, $Ca^{2+}$ ions are removed through solvent exchange using three different solvents: isopropanol (IPA), methanol, and deionized (DI) water: DI water shows the highest binding affinity to $Ca^{2+}$, while IPA shows the lowest[22]. Thus, reassembled chitin nanowires generate different phases in IPA gel, methanol gel, and hydrogel. In IPA and methanol gel states, a nematic liquid crystal (LC) (**N**) phase is generally observed (Fig. 3d-f), while in hydrogel state, it shows a chiral nematic LC (**N***) phase (Fig. 3g-i, for an illustration see the hydrogel panel in Fig. 1b).

*Results and discussion*

The nanostructures of chitin nanowires at each gel state were visualized by POM and cryogenic TEM (Cryo-TEM) (Fig. 3). Considering the long axis of chitin nanowires as director vector, **n**$_{chitin}$ (Fig. 1b), POM can show the orientation of chitin nanowires, i.e. the image is dark when molecules are disordered or **n**$_{chitin}$ is parallel to either of polarizer or analyser. POM with the full-wavelength (530 nm) retardation plate can show the detailed orientation, with the image showing a magenta colour when the sample is isotropic or disordered and blue/yellow domains when **n**$_{chitin}$ is parallel/perpendicular, respectively, to the slow axis of the retardation plate. In the Ca-methanol gel, there are no specific morphologies in POM images (Fig. 3a, b) due to the randomly oriented or disordered chitin nanowires shown in the TEM image (Fig. 3c). POM images of the IPA gel show a well-oriented domain in a large area (~ mm$^2$), in which the **N** phase is found[9]. This is confirmed with TEM (Fig. 3f), which shows one-dimensionally aligned chitin nanowires. The chitin nanowire-hydrogel forms an **N\*** phase, showing the fingerprint patterns in POM images, in which alternating bright and dark lines are present (Fig. 3g). The periodicity of the stripe pattern is given by the helical pitch P/2 (Fig. 1b), giving a pitch P of ~10 to 40 μm with a broad distribution due to the irregular out-of plane arrangement of chitin nanowires. A corresponding TEM image shows a Bouligand-type organization of chitin nanowires (Fig. 3i), which is typically found in the **N\*** phase[23-26]. The counter-clockwise direction of the morphologies means the chitin nanowires assemble into left-handed helical structures in the gel state[12,26,27].

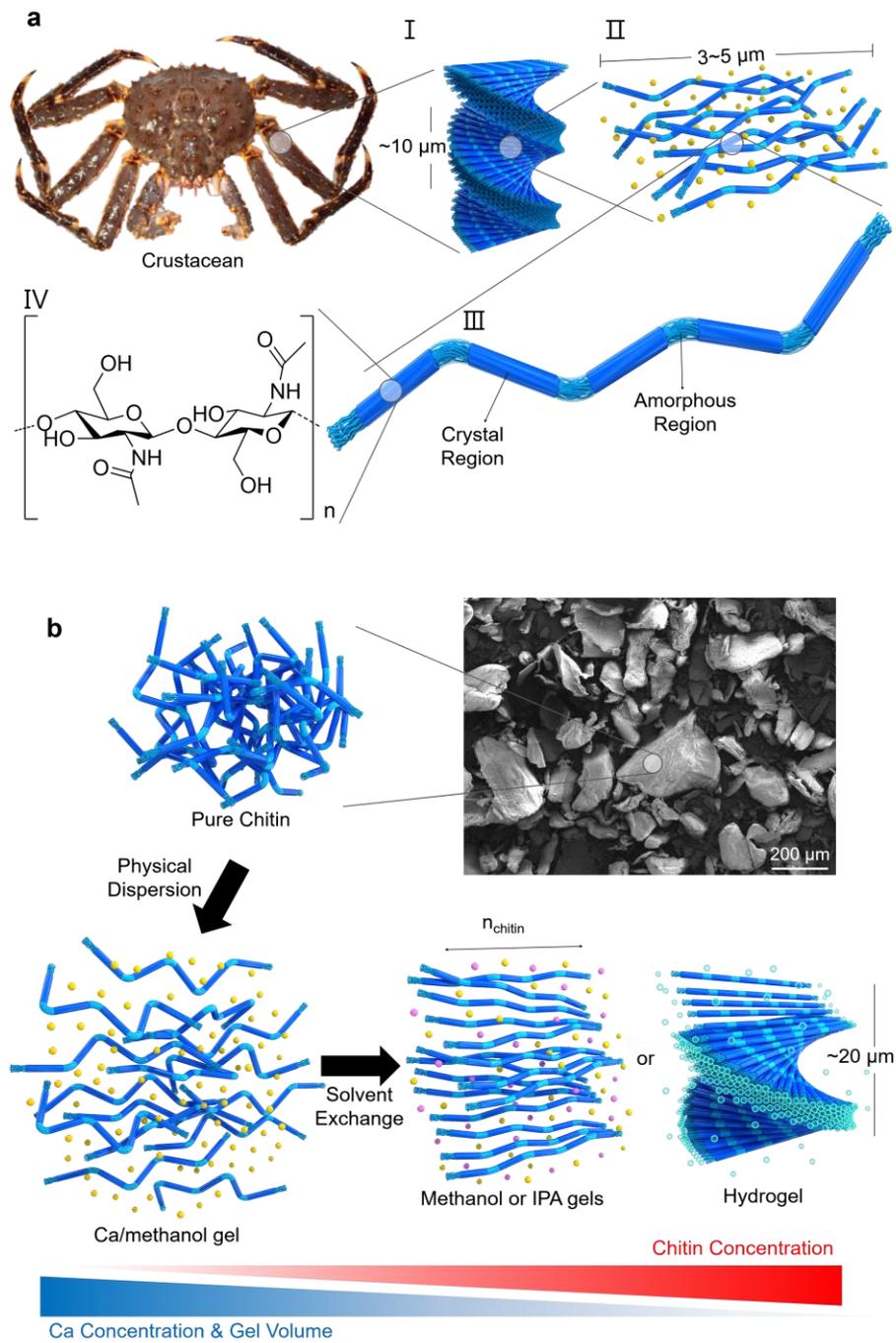

**Figure 1. Hierarchically-ordered chitin microstructure in an arthropod cuticle.** (a) (I) Plywood structure of chitin nanofibrils, (II) chitin nanofibrils in the matrix (CaCO$_3$ or proteins), (III) crystalline and amorphous domains of chitin nanofibril structure, and (IV) chitin structural formula. (b) Calcium-saturated methanol disintegrates chitin nanofibrils with minimal chemical modification, generating a Ca-methanol gel (disordered) (*bottom-left* panel). Ca$^{2+}$ are removed from the Ca-methanol gel by washing with alcohol (methanol or IPA) and DI water, thus generating alcohol gels (methanol gel or IPA gel) in the **N** phase (*bottom-middle* panel) and a hydrogel in the **N*** phase (*bottom-right* panel). The yellow, pink, and blue beads represent three different types of solvent molecules: methanol-solvated Ca$^{2+}$, alcohol (methanol or IPA), and water.

These **N** and **N\*** phases are different from the natural states in terms of the domain size[11,12]. Unlike the previously reported **N\*** phase of highly concentrated nanocellulose in silica[23], our resultant chiral hydrogel shows a finite domain size of 200 to 600 μm (Fig. 3g,h), which may be caused by an energetic penalty of having a large domain size. Here, the charge level of the composite can be divided into 3 classes; i) neutral, when the natural chitin materials that are partially positively charged are mixed with negatively charged matrix materials including ions and proteins;[27] ii) slightly positively charged chitin materials[28] made via natural way with minimal surface deacetylation (Fig. 1b), and iii) strongly positively charged chitin materials made by severe surface deacetylation (Fig. 2). In our resultant **N\*** phase, due to the long-range nature of electrostatic interactions, the energetic penalty can be increased as the domain grows much faster than the energetic gain of having H-bond, which is a short-ranged interaction and thus grows linearly. At a certain size of the domain, both energy loss and gain compensate each other and the growing stops, resulting in finite N* domains. Contrary to this, chitin materials in nature can nucleate and grow to form large or infinite-sized domains without any energetic penalty, despite the charged nature of chitin, since in nature compensating opposite charges are present nearby[27], and charge neutrality is valid locally.

Since chitin has poor solubility in water and alcohol, the solvent exchange in absence of $Ca^{2+}$ results in agglomerated structures. The binding affinity of $Ca^{2+}$ is the highest to water molecule and the lowest to IPA, thus, $Ca^{2+}$ binds less to chitin in water than in IPA. The chitin concentration is the highest for hydrogel and the lowest for IPA gel in an inverse order of the

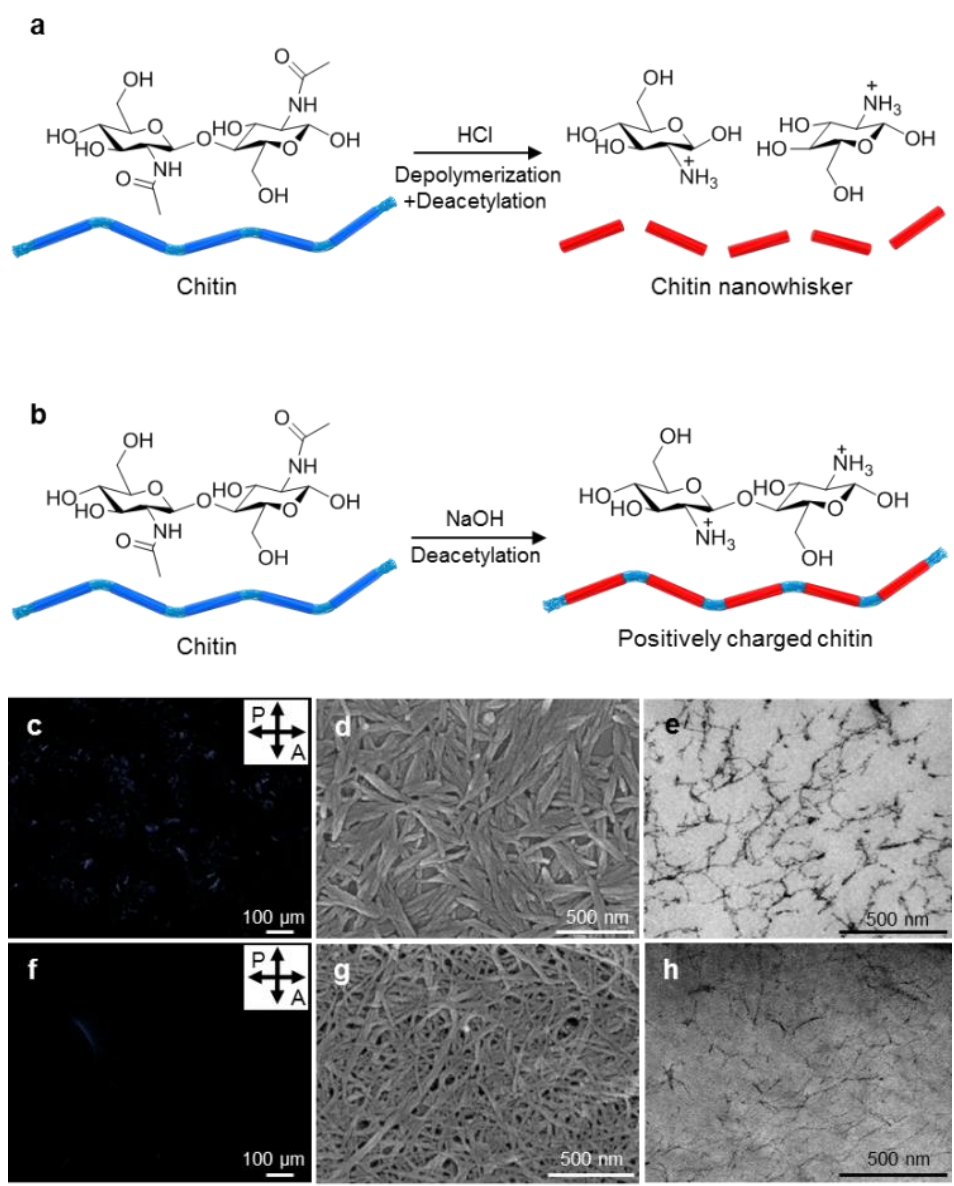

**Figure 2. Chemically modified chitin nanomaterials.** (a) nanowhisker and (b) nanofiber. POM images of (c) nanowhisker and (f) nanofiber solutions (aq). SEM images of (d) dried nanowhisker and (g) dried nanofiber. Cryo-TEM images of (e) nanowhisker and (h) nanofiber solutions (aq).

$Ca^{2+}$ concentration in the equilibrium condition (Fig. S2). Therefore, the amorphous regions of the chitin nanowires are more hydrated and swollen than the rigid crystalline parts in the chitin hydrogel (Fig. S3), not leading to the parallel stacked chitin nanowires but rather to its twisted form (Fig. 3g-i). In this way, the helical pitch of the chitin nanowires in the **N*** phase can be controlled through the water content of the hydrogel (Fig. S4). Evaporation of water generates water flow, which puts hydrodynamic pressure on the nanowires to form more compactly stacked structures. This also reduces the thickness of the hydration shells of amorphous and crystalline domains. Thus, the helical pitch in the **N\*** phase can be decreased. Meanwhile, the thickness of IPA solvation shells in both crystalline and amorphous regions is not substantially different, leading to an **N** phase.

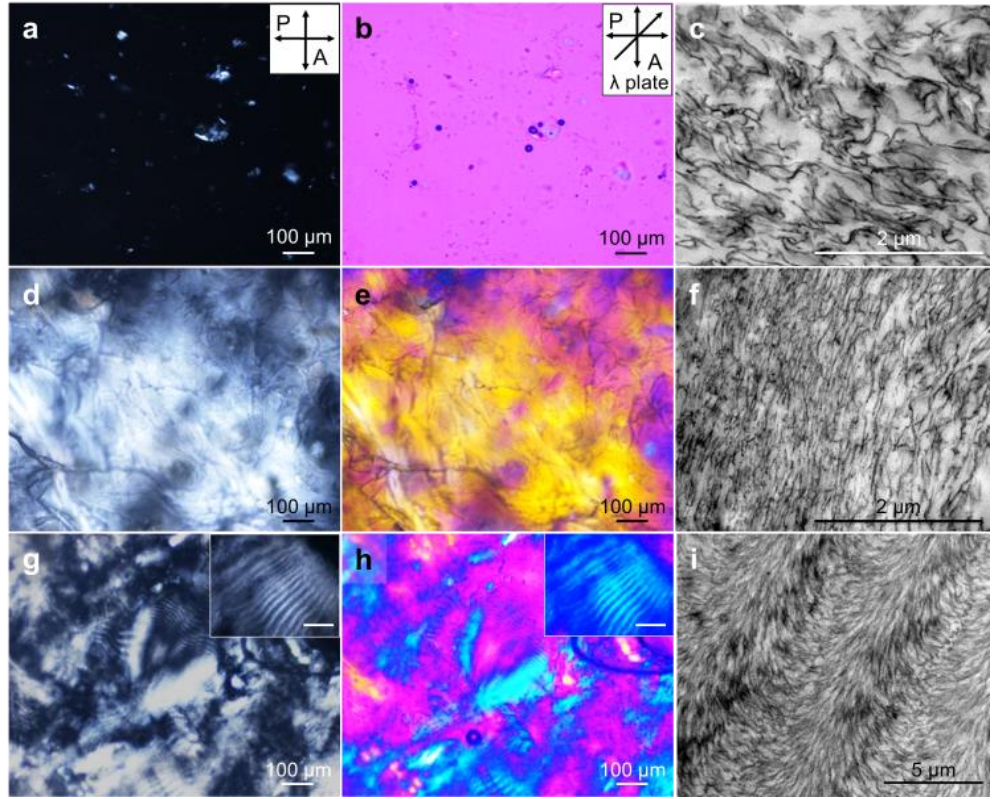

**Figure 3. Morphological change of chitin nanowires by solvent exchange.** POM (left and centre panels) and TEM (right panels) images of Ca-methanol gel (a-c), IPA gel (d-f), and hydrogel (g-i, with an inset scale of 50 µm).

To demonstrate the versatility of our method, the gel-based composite patches were manipulated. The process is simple, and three gel states of chitin nanowires including Ca-methanol gel (disordered), IPA gel (**N**), hydrogel (**N\***) were embedded in an epoxy resin, similar to sample preparation for Cryo-TEM (Fig. 3c, f, i)[29]. Then, we studied the mechanical properties of each type of gel structure. At a fixed chitin content of ~50 g/L in the patches, the Young's modulus ($E$; stiffness) and the toughness (amount of energy stored per unit volume (J/m$^3$) during stretching until failure) were measured for each gel (Fig. 4). Both $E$ and toughness increase as the higher ordered structures are formed (Fig. 4a). However, even though we could not find any substantial difference in $E$ between the **N** and **N\*** phases, the toughness of **N*** is much larger than that of **N** (Fig. 4a, b). In the macroscopic perspective, the finite **N** and **N*** domains are randomly orientated in the patches, meaning that $E$ does not depend much on the hierarchical structures of the chitin nanowires embedded in the patches. During deformation, the **N*** domains redirects the crack direction and prevents crack propagation in the composite[11,12,14,30], giving rise to the notable enhancement in the toughness. It is practically difficult to remove high concentrations of $CaCl_2$ (~670 g/L) from the Ca-methanol using acetone at -80 °C. Therefore, the epoxy resin was not effectively impregnated in mineral-containing chitin nanowires, and the poor interfacial adhesion between the $CaCl_2$ aggregate and the epoxy resin may explain why the composite from the Ca-methanol gel has the weakest mechanical properties.

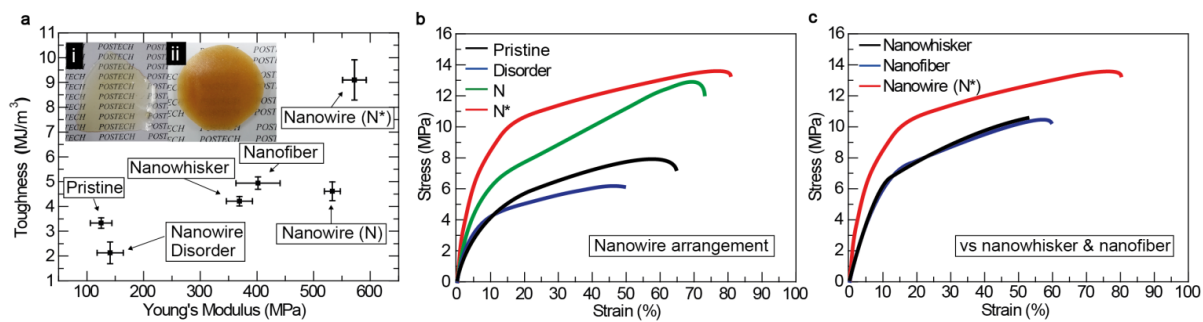

**Figure 4. Mechanical reinforcing effects of chiral nematic chitin nanowires to epoxy resin.** (a) Tensile Young's modulus and toughness of pristine epoxy and chitin epoxy composites. The inset pictures (i) and (ii) show pristine epoxy and chiral nematic chitin epoxy composite films, respectively. (b) Tensile stress-strain curves of epoxy composites with disordered, nematic (N), and chiral nematic (N*) phases of chitin nanowires. (c) Tensile stress-strain curves of chitin nanowhisker and nanofiber epoxy composites are compared.

To confirm the morphological characteristics and the chemical intactness of chitin nanowires, the chitin nanowhiskers and nanofibers were also impregnated in the epoxy resin with the same chitin content of ~50 g/L. Both nanowhisker and nanofiber composites gave much poorer tensile properties in terms of $E$ and toughness when compared to our resultant nanowire patches (Fig. 4c). This shows that the degradation of chitin molecules during the chemical modification softens chitin nanowhiskers and nanofibers, and the resulting high density charges prevent the formation of high ordered structures.

Our chiral nematic chitin nanowires can be applied to the biocompatible composite of polyethylene glycol (PEG) hydrogel (Fig. S5 and Fig. S6). The tensile properties of the PEG hydrogel patches also show that the chitin nanowire based patch has the highest toughness.

*Conclusions*

In summary, we tried to understand and reproduce the chiral plywood structures using chitin nanowires extracted from the crustacean shell via natural way. A chiral nematic arrangement of minimally modified chitin, an unprecedented hierarchical assembly of nanofibril natural polymers in aqueous phase, was obtained. The mechanical properties of the resultant composites demonstrated changes according to the conformation of chiral nanomaterials. Our resultant chiral chitin nanowires can contribute to the understanding of the load bearing systems in nature and potentially enable new biocompatible reinforcing applications.

*Acknowledgments*

This work was supported by the grant from the National Research Foundation (NRF), funded by the Korean Government (MSIP) (2014R1A2A2A01006724, C1ABA001-2011-0029960, 2015R1A1A1A05000986 and 2012M3A7B4049802). This work was also supported by the

Marine Biomaterials Research Centre grant funded by the Ministry of Oceans and Fisheries, Korea.

*Author Contributions*

D.X.O., D.S.H and D.K.Y. designed research; D.X.O., Y.J.C., H.L.N. and H.H.J. performed research; D.X.O., Y.J.C, Y.S.J., D.S.H and D.K.Y. analysed data; D.X.O., Y.S.J., D.S.H and D.K.Y. wrote the paper.

*Author information*

Reprints and permissions information is available at www.nature.com/reprints. The authors declare no competing financial interests. Correspondence and request for materials should be addressed to DKY and DSH. (nandk@kaist.ac.kr and dshwang@postech.ac.kr)

# Supplementary Information

**Chiral nematic self-assembly of minimally surface damaged chitin nanofibrils and its load bearing functions**

*Dongyeop X. Oh[1,2,+], Yun Jeong Cha[3,+], Lihn H. Nguyen[4], Hwaheon Je[4], Yong Seok Jho[5,6], Dong Soo Hwang[1,4,7,*], & Dong Ki Yoon[3,*]*

[1]POSTECH Ocean Science and Technology, Pohang University of Science and Technology (POSTECH), Pohang 790-784, Republic of Korea.

[2]Research Centre for Industrial Chemical Biotechnology, Korea Research Institute of Chemical Technology, Ulsan 44429, Republic of Korea

[3]Graduate School of Nanoscience and Technology and KINC, KAIST, Daejeon 305-701, Republic of Korea

[4]Division of Integrative Biosciences and Biotechnology, Pohang University of Science and Technology (POSTECH), Pohang 790-784, Republic of Korea

[5]Department of Physics, Pohang University of Science and Technology, [6]Asia Pacific Center for Theoretical Physics, Pohang 790-784, Republic of Korea

[7]School of Environmental Science and Engineering, Pohang University of Science and Technology (POSTECH), Pohang 790-784, Republic of Korea

[+] These authors contributed equally to this work.

*Correspondence should be addressed to Dong Soo Hwang (e-mail: dshwang@postech.ac.kr) & Dong Ki Yoon. (e-mail: nandk@kaist.ac.kr)

**Methods**

*Materials*

Calcium chloride di-hydrate ($CaCl_2 \cdot H_2O$) and methanol were purchased from Junsei Chemical (Japan). Chitin powder, epoxy embedding medium kit, and polyethylene glycol diacrylate (PEGDA; 700 g/mol) were purchased from Sigma-Aldrich. IPA and 2-hydroxy-4'-(2-hydroxyethoxy)-2 methylpropiophenone (photo-initiator; PI) were purchased from TCI Chemical (Japan). All aqueous solutions were prepared using deionized (DI) water.

*Physical disintegration of chitin to chitin nanowire: Ca-methanol gel*

Ca-methanol was prepared as follows[10]. Calcium chloride di-hydrate (428 g) and DI water (30 ml) were added to methanol (470 ml). The mixture was refluxed at 150 °C for 6 h. To produce Ca-methanol gel, chitin powder (8 g) was added to the Ca-methanol (400 ml), then the mixture was refluxed in at 150 °C for 6 h and cooled down at room temperature.

*Solvent exchange of Ca-methanol gel to produce hydrogel, methanol gel, and IPA gel*

$Ca^{2+}$ ions were removed from the Ca-methanol gel via solvent-exchange with three different solvents: DI water, methanol, and IPA[10]. The Ca-methanol gel was immersed in excess volume of each exchange solvent with vigorous stirring for 12 h, then filtrated by nylon cloth to collect the chitin gel precipitate. The chitin gel precipitates were centrifuged at 8000 rpm for 15 min to remove excess of solvent. Consequently, hydrogel (=chiral chitin nanowire hydrogel), methanol gel, and IPA gel were obtained.

*Chitin and calcium concentration*

Chitin concentration (weight per volume; g/L) of each gel was evaluated as follows. Ten millilitre of each gel was incubated in EDTA buffer at 37 °C for 48 h with shaking and vigorously washed with DI water. The insoluble part was dried at 100 °C in a vacuum oven for 24 h. And then, the dry weight of the insoluble part was measured.

Calcium concentration (weight per volume; g/L) of each gel was evaluated as follows. Ten millilitre of each gel was dried at 100 °C in a vacuum oven for 24 h. And then, the weight of the dried sample was measured. Calcium weight in the dry sample was investigated using an inductively coupled plasma atomic emission spectroscopy (ICP-AES) on an IRIS Advantage 1000.

*Preparation of chitin nanowhisker and nanofiber*

Chitin nanowhisker was prepared as follows.[17] Chitin powder (20 g) was added into 3 N HCl (400 ml). And then, the mixture was refluxed at 150 °C for 6 h. The solution was purified 3 times with DI water using a series of dilution-centrifugation-decantation process. The suspension was dialyzed overnight in DI water, then lyophilized. The lyophilized chitin nanowhisker was re-dispersed in 50 mM acetate buffer (pH 4) via ultra-sonication. Chitin nanofiber was prepared as follows[16-18]. Chitin powder (20 g) was suspended in 20 wt% NaOH (*aq*) (400 ml), then refluxed at 150 °C for 6 h. The wet sample was vigorously washed with DI water, then dispersed in DI water to be ~1.3 wt %. Drops of acetic acid was added into the chitin suspension to adjust the pH value to ~4. For disintegration, the chitin suspension passed through a high performance grinder (MKCA6−3; Masuko Sangyo Co., Ltd.) with the rotation speed of 1500 rpm and a clearance gauge of −1.5. The chitin nanofiber suspension was dialyzed overnight in DI water, then lyophilized. The lyophilized chitin nanofiber was re-dispersed in 50 mM acetate buffer (pH 4) by ultra-sonication.

*Cryo-TEM*

Each sample was placed on gold-plated membrane carriers[29, 31]. The sample-containing carrier was loaded into a Leica EM HPM100 (Germany) and rapidly frozen with liquid nitrogen under high pressure of 2100 bar. The sample was freeze-substituted with 0.2% uranyl acetate-containing dry acetone at −80 °C for 3 days. The sample was warmed from −80 °C to −20 °C, immersed in pure acetone for 3 day, infiltrated with acetone/epoxy resin precursor (epoxy embedding medium kit) mixtures (4:1 for 75 min and 2:1 for 2 h) at −20 °C, and infiltrated with the pure epoxy resin precursor for 12 h at room temperature. The sample was cured at 70 °C for 24 h. The resin was cut into 200 nm-sections using Leica EM UC7 (Germany) and applied onto copper grids. All TEM images were recorded using a JEOL JEM 1011 transmission electron microscope. The resin precursor contains epoxy embedding medium (20 ml), 2-dodecenylsuccinic anhydride (9 ml), methyl nadic anhydride (12 ml), and 2,4,6-tris(dimethylaminomethyl)phenol (1 ml).

*Equilibrium water content (EWC) and equilibrium alcohol content (EAC)*

To characterize the degree of water and alcohol absorption, the EWC and EAC were measured[32]. Dry chitin (or regenerated chitin) was weighed to an accuracy of $10^{-4}$ g and immersed in water (or isopropanol) for 1 day. And then, the immersed film was weighed again after removal of excess liquid. EWC or EAC were calculated using the following equation:

$$\frac{W_t - W_0}{W_t} X\ 100$$

$W_0$ is the weight of the sample before immersion and $W_t$ is the weight of the sample after immersion at time t when the weight of the sample reaches the equilibrium state.

The regenerated chitin was prepared as follows. One gram of chitin powder was dissolved in nine gram of 1-ethyl-3-methylimidazolium acetate ([$C_2$mim] [OAc], an ionic liquid) at 100 °C for 6 h with stirring. The solution was cooled down in ambient conditions for 1 day to form a gel. The gel was soaked in excess of water for 6 h and, then washed with distilled water.

*Chitin nanowire-embedded epoxy composites*

Ca-methanol gel, IPA gel, and hydrogel were rapidly frozen with liquid nitrogen[29, 31]. Each sample was freeze-substituted with pure acetone at −80 °C for 3 days and at −20 °C for 3 days. The sample was infiltrated with acetone-epoxy resin precursor mixtures (4:1 for 1 day and 2:1 for 1 day) at −20 °C, and infiltrated with the pure epoxy resin precursor at room temperature for 12 h. The chitin/epoxy resin precursor suspension was centrifuged at 8000 rpm for 20 min. The desire amount of the epoxy resin precursor was added into the precipitate-containing bottle to adjust chitin contration. And then, the mixture was homogenized with a domestic blender and an ultra-sonicator. The suspension was cured at 70 °C for 24 h. The epoxy resin precursor contains epoxy embedding medium (20 ml), 2-dodecenylsuccinic anhydride (22 ml), methylnadic anhydride (2 ml), and 2,4,6-tris(dimethylaminomethyl)phenol (0.6 ml).

*Nanofiber- and nanowhisker-embedded epoxy composites*

Nanofiber (or nanowhisker)-dispersed solution at the chitin concentration of ~20 g/L was rapidly frozen with liquid nitrogen. The sample was freeze-substituted with pure acetone at −80 °C for 3 days and at −20 °C for 3 days. The sample was infiltrated with acetone-epoxy resin precursor mixtures (4:1 for 1 day and 2:1 for 1 day) at −20 °C, and infiltrated with the pure epoxy resin precursor at room temperature for 12 h. The chitin/epoxy resin precursor suspension was centrifuged at 800 rpm for 20 min. The desire amount of the epoxy resin precursor was added into the precipitate-containing bottle to adjust chitin concentration. And then, the mixture was homogenized with a domestic blender and an ultra-sonicator. The suspension was cured at 70 °C for 24 h. The epoxy resin precursor contains epoxy embedding medium (20 ml), 2-dodecenylsuccinic anhydride (22 ml), methylnadic anhydride (2 ml), and 2,4,6-tris(dimethylaminomethyl)phenol (0.6 ml).

*Chiral chitin nanowire-embedded PEG hydrogel composites*

Chiral chitin nanowire-embedded polyethylene glycol (PEG) hydrogels were prepared[33]. PEGDA is a precursor of PEG hydrogel, and PI is an UV-responsive cross-liking agent of the PEGDA. The chiral chitin nanowire hydrogel has chitin concentration of ~60 g/L. Desired amount of water, PEGDA, and PI were added into the chiral chitin nanowire hydrogel: PEGDA (150 g/L), PI (1.5 g/L), and chitin concentration (0, 5, 10, or 20 g/L). And then, the mixtures were vigorously homogenized using a domestic blender and an ultra-sonicator. The homogenized suspensions were incubated at 4 °C for 3 days for PEGDA infiltration, then cured by 15 min UV irradiation.

*Nanowhisker- and nanofiber- embedded PEG hydrogel composites*

Nanowhisker and nanofiber-embedded PEG hydrogels were prepared as follows[33]. Desired amount of PEGDA, PI, and lyophilized chitin nanomaterial were added into 50 mM acetic acid buffer: PEGDA (150 g/L), PI (1.5 g/L), and chitin concentration (20 g/L). The mixtures were vigorously homogenized using a domestic blender and an ultra-sonicator. The homogenized suspensions were incubated at 4 °C for 3 days for PEGDA infiltration, then cured by 15 min UV irradiation.

*XRD*

The crystal structure of chitin and chiral chitin nanowire was studied using wide-angle X-ray diffraction experiment. The experiment was conducted on an X-ray diffractometer (XRD) (D/MAX-2500/PC, Rigaku, Japan), in which a 40 kV/100 mA Ni-filtered Cu Kα radiation was employed. The WRD patterns were recorded in the region of a scattering angle of 5° to 40° with a scanning speed of 4°/min.

*Tensile tests*

    1) Chitin/epoxy composites

The tensile properties of the chitin nanowire/epoxy composites were measured on a universal tensile tester (UTS, Instron, Norwood, UK). All samples were cut into the following dog-bone shape.

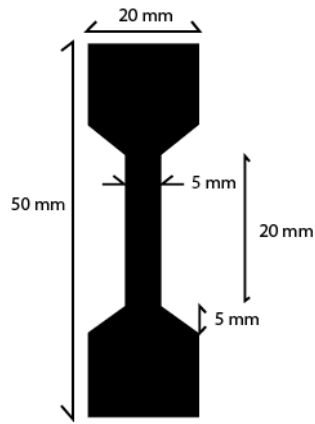

Each sample was clamped onto the grips and loaded with a constant strain rate of 5 mm/min till failure. To calculate the tensile stress (σ, MPa, load per unit cross-sectional area) on the films, thickness and width of the films were determined using a micrometer before the test. In the strain-stress curve, Young's modulus was determined as the initial slope of the curve at 0.2% strain and toughness was determined as the area under the curve until its failure.

    2) Chitin PEG hydrogel composites

The tensile properties of the chitin/PEG hydrogel composites were measured on a universal tensile tester (UTS, Instron, Norwood, UK). All PEG hydrogels were cut into 30 mm × 5 mm × 4 mm cuboid shape. Each sample was clamped onto the grips with 10 mm distance and loaded with a constant strain rate of 5 mm/min till failure. To calculate the tensile stress (σ, MPa, load per unit cross-sectional area) on the films, thickness and width of the films were determined using a micrometer before the test. In the strain-stress curve, Young's modulus was determined as the initial slope of the curve at 0.2% strain and toughness was determined as the area under the curve until its failure.

**Supplementary text**

*EWC and EAC of pure chitin and regenerated chitin*

Chitin is neither meltable nor soluble in most solvents. Only solvents, such as NaOH/urea (*aq*) and ionic liquids, have been found to successfully dissolve chitin at molecular level[32]. Re-solidified chitin from those of chitin solutions is called "regenerated chitin", which have low crystallinity, in other word high content of amorphous region. Thus, the EWC data (Fig.S4) suggests that the crystalline region of chitin is less hydrated than the amorphous domain of chitin. However, the difference between EAC values of pure chitin and regenerated chitin was much lower than the difference between EWC values.

*Tensile properties of the chitin nanowire-embedded PEG hydrogel.*

The chiral chitin nanowire was impregnated into the PEG hydrogel (Fig. S5). To investigate how the morphology of the chitin nanomaterials affects their reinforcing property, nanowhisker and chitin nanofiber were also impregnated into the PEGDA[33]. At the same chitin concentration (20 g/L), there's no remarkable difference in *E* values for all of three types of chitin nanomaterials-embedded PEG hydrogels, while the toughness is much varied with the chiral arrangement of the building blocks (Fig. S6). Thus, the toughness relies on the conformation of the chitin building blocks, related with the enduring energy against the rupturing. In details, the chiral nematic structure of chitin nanowires has the multi-hierarchal arrangement of the chiral building blocks, which prevent the formation of micro-cracks and macro-crack propagation[5, 12, 14, 30]. And reasonably the stiffness and toughness of the hydrogels increased with the chiral chitin nanowire concentration.

**Supplementary figures**

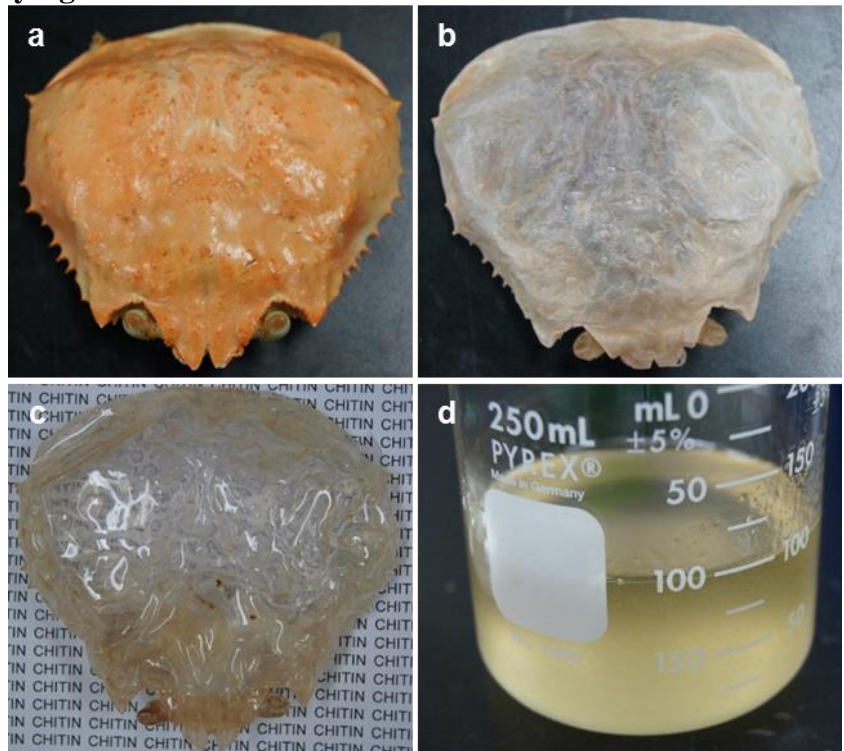

**Figure S1.** (a) A crab shell and (b) purified chitin skeleton from the crab shell. The chitin skeleton was immersed in Ca-methanol at (c) 70 ℃ and (d) 150 ℃.

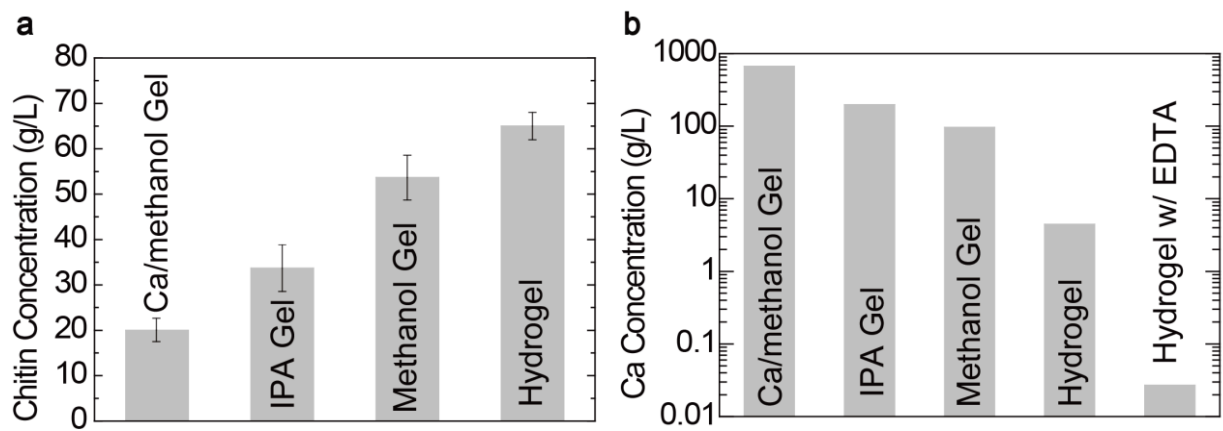

**Figure S2.** (a) Chitin concentration (weight per volume, g/L) of Ca/methanol gel, IPA gel, methanol gel, and hydrogel. Each value and error bar represents the mean of quadruplicate samples and its standard deviation. (b) Calcium concentration (weight per volume, g/L) of Ca-methanol, IPA gel, methanol gel, hydrogel, and EDTA-treated hydrogel from the inductively coupled plasma spectroscopy (ICP).

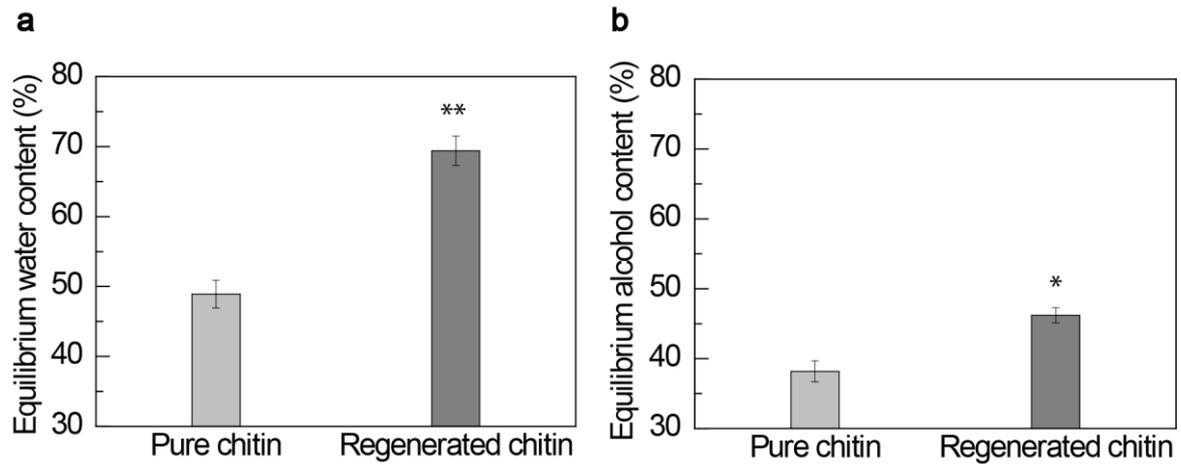

**Figure S3**. Equilibrium water content (EWC) and equilibrium alcohol content (EAC) of pure chitin and regenerated chitin after immersion in solvents, (a) water and (b) isopropanol (IPA) respectively, for 1 day. [n=3; mean ± the standard error of the mean; *, P<0.05; **, P<0.005)]

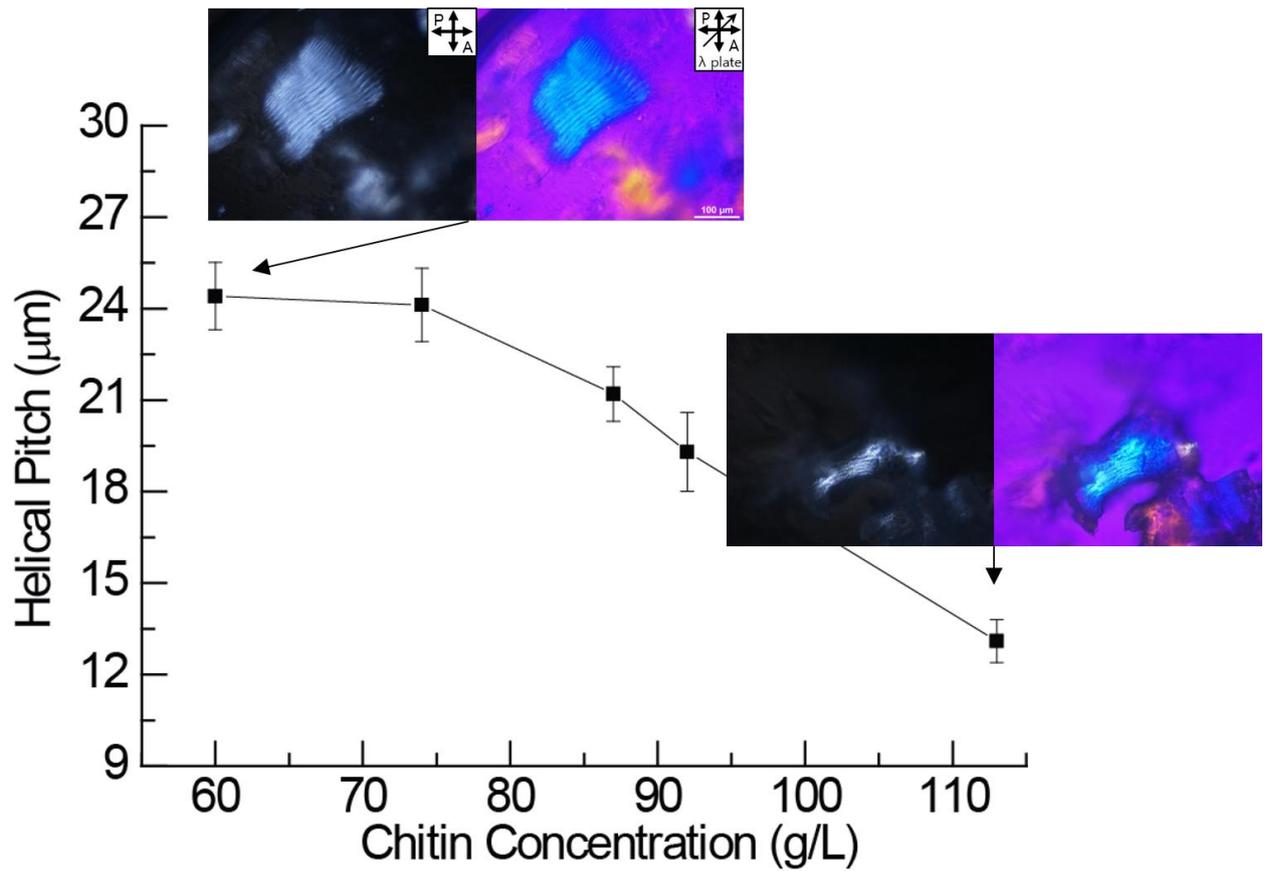

**Figure S4**. Helical pitch of chiral nematic chitin nanowire hydrogel depending on chitin concentration during water evaporation. Each value and error bar represents the mean of triplicate samples and its standard deviation.

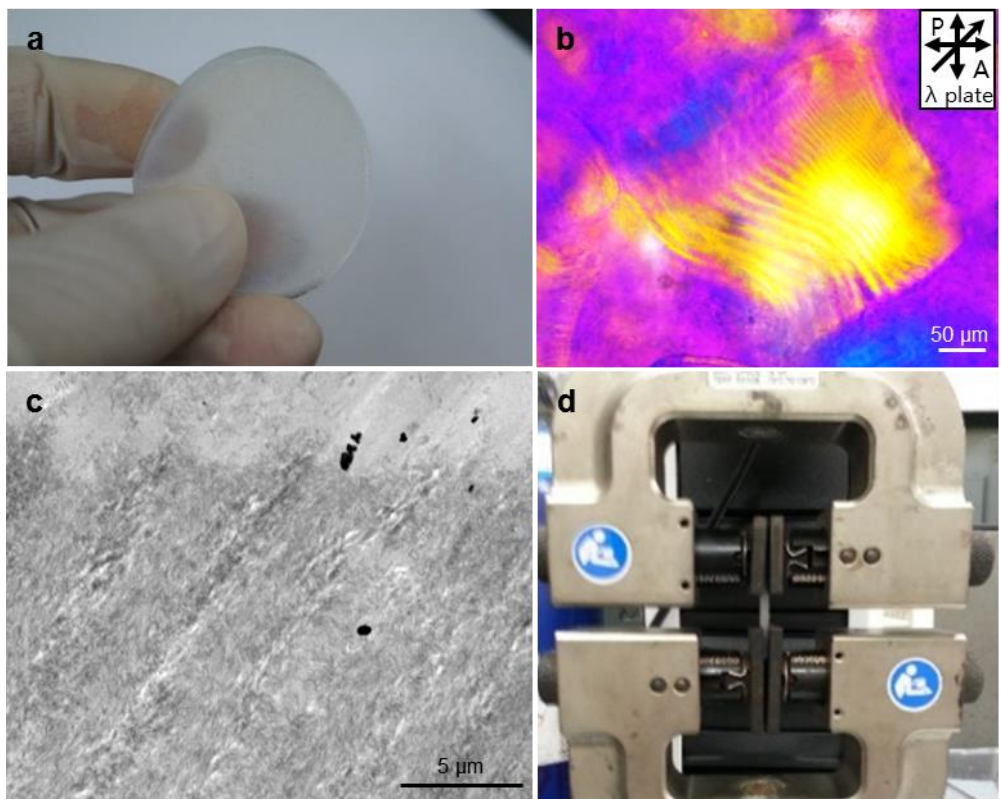

**Figure S5**. (a) Optical camera, (b) polarized optical microscopy, and (c) Cryo-TEM images of chiral nematic nanowire PEG hydrogel patch. (d) Experimental set-up for the tensile tests.

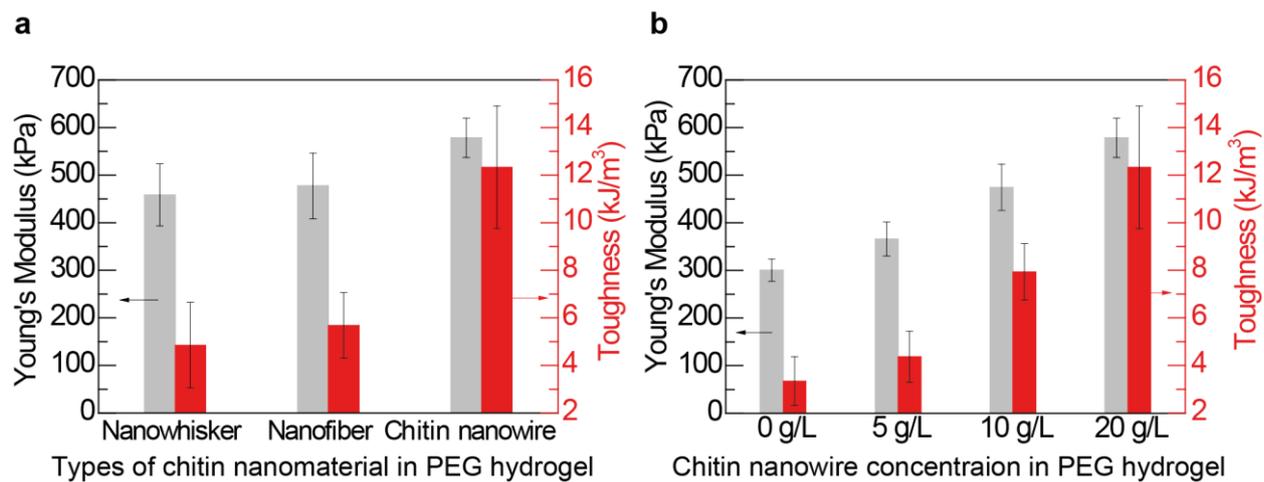

**Figure S6.** Tensile Young's modulus ($E$) and toughness of chitin nanomaterial-embedded PEG hydrogel patches depends on (a) types of chitin nanomaterials at the same chitin concentration (~20 g/L) and (b) chiral chitin nanowire concentration in the PEG hydrogel. Each value and error bar represents the mean of quadruplicate samples and its standard deviation.

**Supplementary References**